\documentclass[runningheads,a4paper]{llncs}

\usepackage{amssymb}
\usepackage{graphicx}
\usepackage{natbib}
\usepackage{url}
\newcommand{\keywords}[1]{\par\addvspace\baselineskip
\noindent\keywordname\enspace\ignorespaces#1}

\begin{document}

\mainmatter  

\title{Music Generation with Variational Recurrent Autoencoder Supported by History}

\titlerunning{Music generation with VRASH}

%
%
\author{Ivan P. Yamshchikov\inst{1} \and Alexey Tikhonov\inst{2}}
%
\authorrunning{Ivan P. Yamshchikov and Alexey Tikhonov}

\institute{Max Planck Institute for Mathematics in the Sciences, Leizpig  \email{ivan@yamshchikov.info} \and Yandex, Berlin \email{altsoph@gmail.com}}

%
%

\maketitle

\begin{abstract}
A new architecture of an artificial neural network that helps to generate longer melodic patterns is introduced alongside with methods for post-generation filtering. The proposed approach called variational autoencoder supported by history is based on a recurrent highway gated network combined with a variational autoencoder. Combination of this architecture with filtering heuristics allows generating pseudo-live acoustically pleasing and melodically diverse music.

\end{abstract}

\section{Introduction}\label{introduction}
 
A rapid progress of artificial neural networks is gradually erasing the border between the arts and the sciences. A significant number of results demonstrate how the areas that were previously regarded as entirely human due to the creative or intuitive nature of the tasks transform and give space for the algorithmic approaches \cite{Ano18}. Music is one of these areas. Indeed, there was a number of attempts to automate the process of music composition long before the artificial neural networks era. A well-developed theory of music inspired a number of heuristic approaches to automated music composition. The earliest idea that we know of dates as far back as the nineteenth century, see \cite{lovelace}. In the middle of the twentieth century, a Markov-chain approach for music composition was developed in \cite{hiller}. However, as it was shown in \cite{lin} music, as well as some other types of human-generated discrete time series, tends to have long-distance dependencies that can not be captured with a Markov-chain based model. Recurrent neural networks (RNNs), on the other hand, better process data-series with longer internal dependencies \cite{sundermeyer}, such as sequences of notes in a tune \cite{bl}. Indeed, a variety of different recurrent neural networks such as hierarchical RNN, gated RNN, Long-Short Term Memory (LSTM) network, Recurrent Highway Network etc. were successfully used for music generation in \cite{chu}, \cite{colombo}, \cite{johnson}, \cite{choi}, \cite{zilly} or  \cite{magenta}. 

The similarity between problem setup for note-by-note music generation and word-by-word generation of text makes it reasonable to overview the methods that proved themselves useful for generative natural language processing tasks. We would like to focus on a variational recurrent autoencoder (VAE) proposed in \cite{bowman}, \cite{semeniuta}. VAE is making assumptions concerning the distribution of latent variables and applies variational approach for latent representation learning. This yields an additional loss component and a specific training algorithm called Stochastic Gradient Variational Bayes (SGVB) \cite{rezende}, \cite{kingma}. Thus a generative VAE obtains examples similar to the ones drawn from the input data distribution. It also gives a significant control over the parameters of the generated output, see \cite{larsen}, \cite{yan}. This theoretically opens the door for controlled music output and makes the idea to apply VAE-based method to music generation very inviting.

Despite a number of advantages mentioned above, artificial neural networks also have a well-known problem when applied to music or language generation. A significant percentage of generated sequences, despite their statistical similarity to the training data, are flagged as wrong, boring or inconsistent, when reviewed by human peers. This hinders broader adoption of neural networks in this areas. The contribution of this paper is two-fold: (1) we suggest a new architecture for algorithmic composition of monotonic music called Variational Recurrent Autoencoder Supported by History (VRASH) and (2) we demonstrate that when paired with a simple filtering heuristics VRASH can generate pseudo-live acoustically pleasing and melodically diverse music.

\section{Music representation and data}

Four gigabytes of midi-files that included songs of different epochs and genres formed a proprietary dataset that was used for the experiments. The data was already available but needed a significant preprocessing. Since one midi-file can contain several tracks with meaningful information yet can have some tracks of little importance, the files were split into separate tracks. Certain normalization of the data is often needed to facilitate learning.  The following normalization procedures were applied to every track individually. Each note in midi file is defined with several parameters such as pitch, length, strength plus the parameters of the track (e.g. the instrument that is playing the note) and the parameters of the file (such as tempo). Despite the fact that nuancing is playing an important role in musical compositions the strengths of the notes was omitted. This particular work was focused on the melodic patterns determined by the pitches and temporal parameters of the notes and pauses in between. The median pitch of every track was transposed to the 4th octave. The pauses throughout the dataset were also normalized as follows. For each track a median pause was calculated. It is only to be expected that absolute majority of the pauses in the track are equal to the median pause multiplied with a rational coefficient (1/2 and 3/2 being especially popular for the majority of the tracks). The tracks with more than eleven different values of the pauses were filtered out. Generally, temporal normalization of the midi files might be rather challenging but the pause filtering trick described above allows to normalize the obtained tracks using the value of the median pause. Finally to prevent the model from possible overfitting and to make the input diverse enough the tracks with exceedingly small entropy were also excluded from the training data. Since the tracks are generated on a note-by-note basis an exceeding amount of tracks with low pitch entropy (say, house bass-line with the same note repeating itself throughout the whole track) would drastically decrease the quality of the output. Final dataset consisted of 15+ thousand normalized tracks and was used for further training. 
 
A concatenated note embedding was constructed for every note in a track. This embedding included the pitch of the note, its octave and a delay that corresponded to the length of the note as well as meta-information of a given MIDI track.

\section{Architecture}

\begin{figure}[htpb]
\begin{center}
\includegraphics[width=\linewidth]{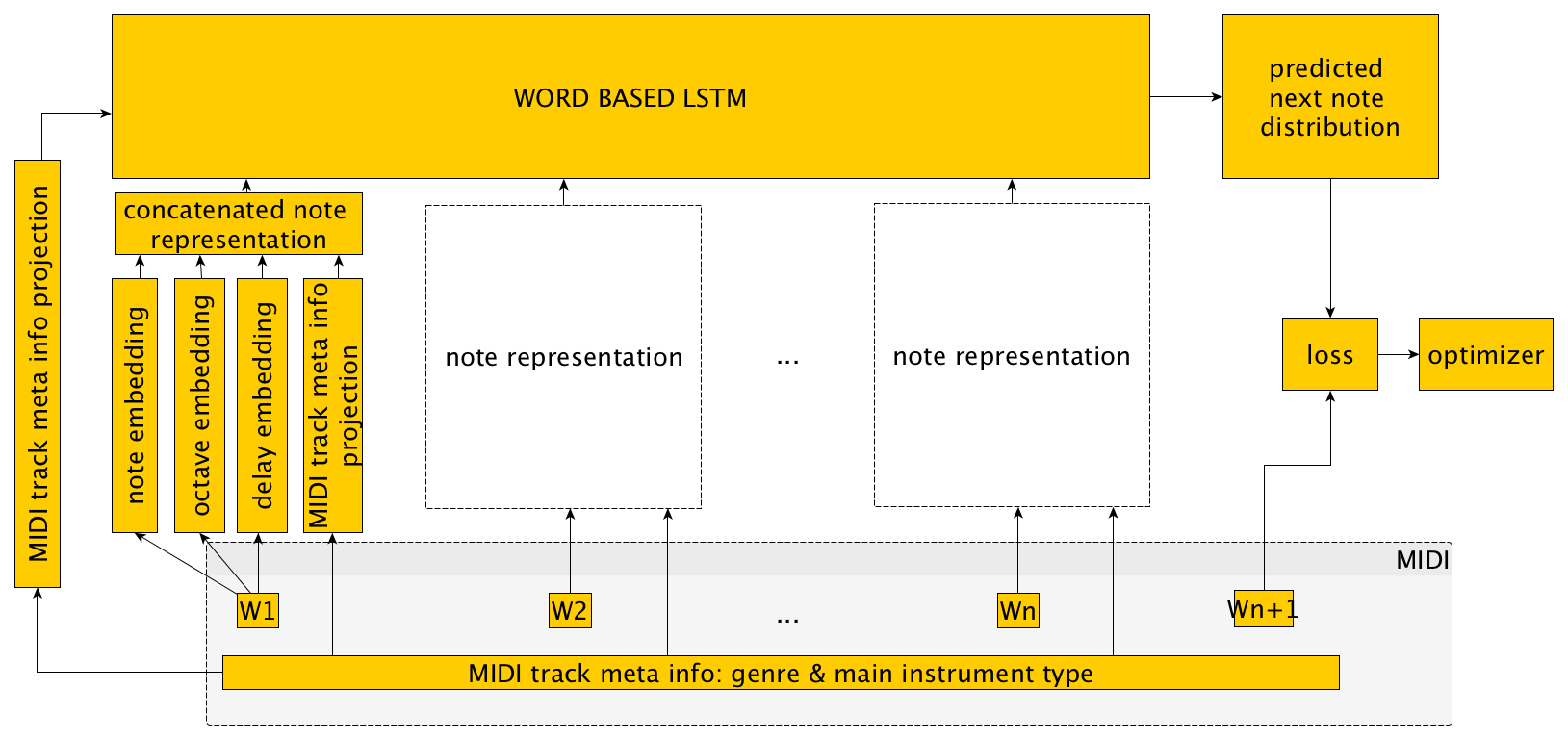} 
\caption{\emph{Language model scheme for music generation.}}
\label{fig:arcLM}
\end{center}
\end{figure}

For the task of music generation we have trained thee different architectures. The baseline for such tasks is usually a classical language model (LM) shown on Figure \ref{fig:arcLM}.
 
\begin{figure}[htpb]
\begin{center}
\includegraphics[width=\linewidth]{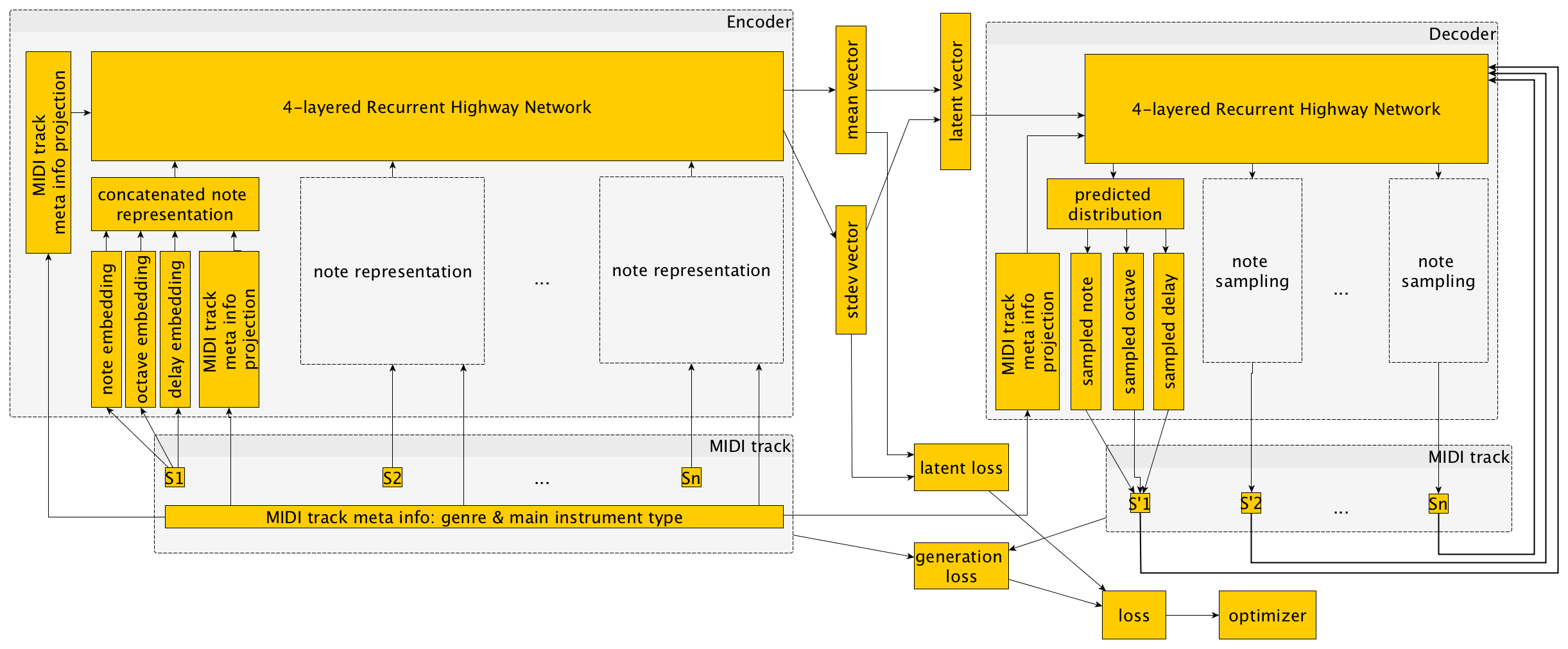} 
\caption{\emph{Variational autoencoder scheme for music generation. Bottleneck between decoder and encoder aims to compress the macrostructure of the melody effectively and obtain a diverse melody with a human-like macrostructure. A variational bayesian noise highlighted with light yellow colour.}}
\label{fig:arcVAE}
\end{center}
\end{figure}

 Variational autoencoder was originally proposed for the tasks of text generation in \cite{bowman}, \cite{semeniuta}. Figure \ref{fig:arcVAE} demonstrates this architecture in application to music generation.
 
\begin{figure}[htpb]
\begin{center}
\includegraphics[width=\linewidth]{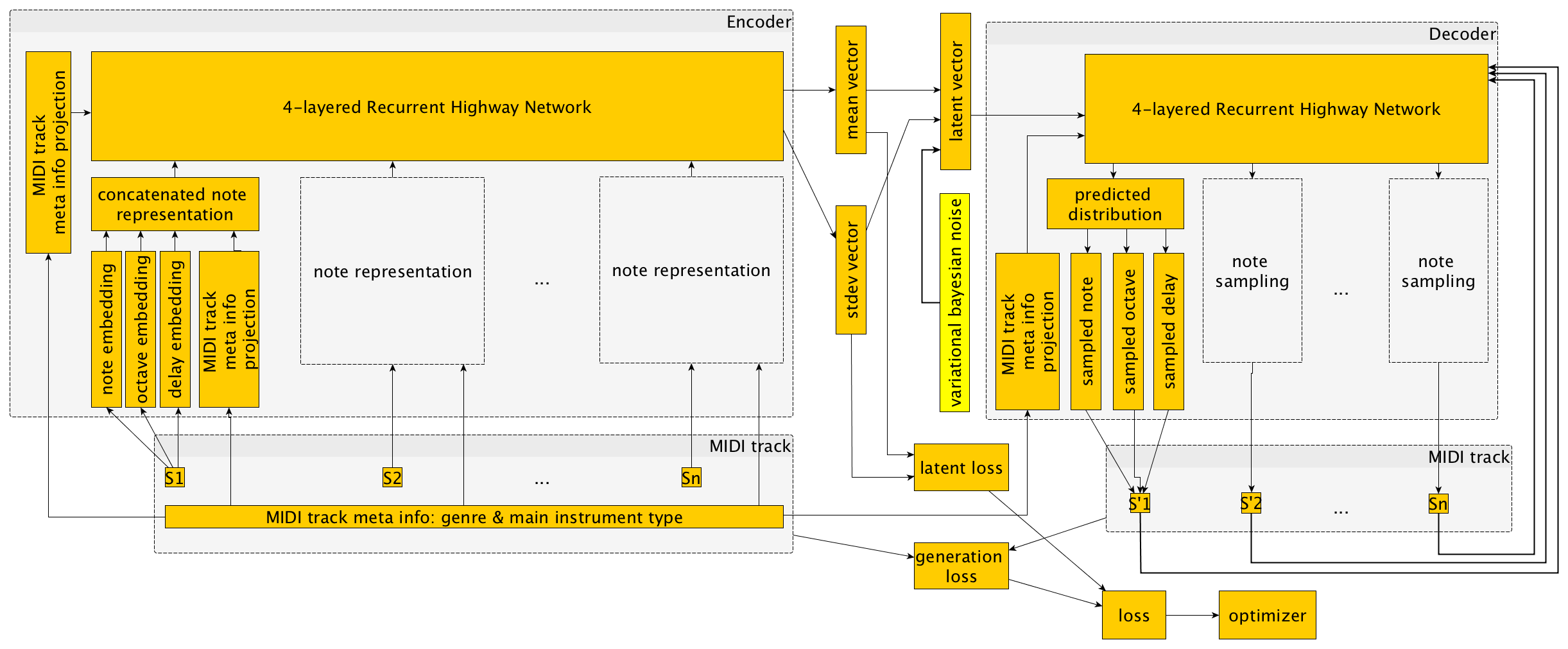} 
\caption{\emph{Variational Recurrent Autoencoder Supported by History (VRASH) scheme for music generation. Previously generated notes are used for the generation of further notes.}}
\label{fig:arcVRASH}
\end{center}
\end{figure}

 In contrast with classical Variational Autoencoder a Variational Recurrent Autoencoder Supported by History shown in Figure \ref{fig:arcVRASH} uses previous outputs as additional inputs. VRASH "listens" to the notes that it has composed already and uses them as additional "historic" input. To our knowledge this is the first application of such generative approach to the generation of music rather than text.
 
The support by history partially addresses the issue of slow mutual information decline and that seems to be typical for natural discrete sequences such as natural language, notes in a composition or even genes in a human genome, as \cite{lin} state. Let us now look at this issue a little closer. The following definitions of mutual information $I$ between two random variables $X$ and $Y$ are equivalent

\begin{eqnarray}
I(X, Y) &\equiv& S(X) + S(Y) - S(X,Y) = D(p(XY)\| p(X)p(Y)) \label{eq:mi} \\
	&=& \Big< log_{2}\frac{P(x,y)}{P(x)P(y)} \Big> = \sum_{x,y} P(x,y) log_{2} \frac{P(x,y)}{P(x)P(y)}, \nonumber 
\end{eqnarray}
 where $S = < -log_{2} P >$ is a Shannon entropy measured in bits, see \cite{Shannon}, and $D$ is the Kullback-Leibler divergence, see \cite{KL}. Indeed, \cite{lin} show that in a number of natural datasets mutual information between such tokens declines relatively slowly. VRASH addresses this problem specifically, trying to compensate for slow mutual information decline with the {\em history support} mechanism. Contrary to the approach proposed in \cite{Roberts}, where network generate short loops and then connects them in longer patterns, thus providing a possibility to control melodic variation and regularity, we focus on the whole-track melody generation. Let us now describe the experimental results obtained.
 
 \section{Experiments and discussion}

Before we discuss the proposed architectures we need to make the following remark. It is still not clear how one could compare the results of generative algorithms that work in the area of arts. Indeed, since music, literature, cinema etc. are intrinsically subjective it is rather hard to approach them with a certain rigorous metric. A majority of approaches is usually based on the peer-review systems where the amount of human peers can significantly vary. For example, in \cite{huang} the authors refer to 26 peers subjective opinions, whereas in \cite{hadjeres} more than 12 hundred peer responses are analyzed. Such collaborative approach based on individual subjective assessment could be used to characterize the quality of the output but is typically costly and hardly can be used to obtain scalable results. The amount of peers needed to compare several different architectures and obtain rigorous quantitative differences between them drastically exceeds the ambition of this particular research. Keeping this remarks in mind we would further discuss possible objective metrics that are standardly used for comparisons of generative models and suggest a simple yet useful workaround for quality assessment. 

\begin{figure}[h]
\begin{center}
\includegraphics[width=\linewidth]{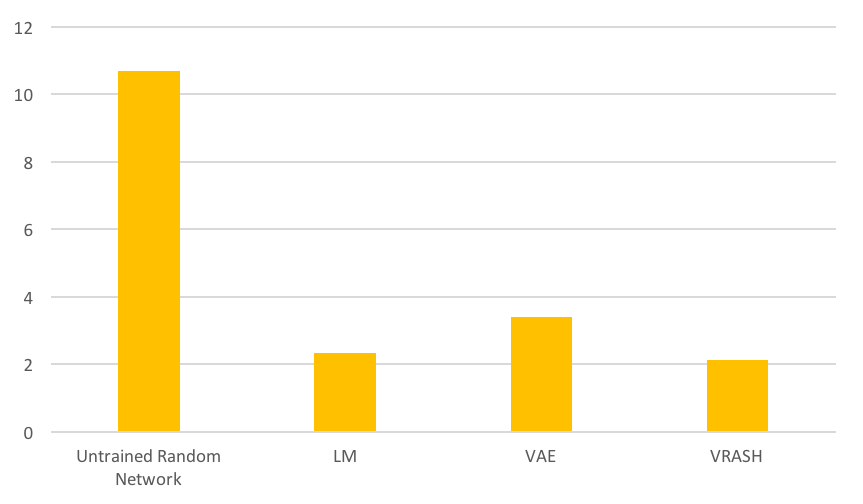} 
\caption{\emph{Cross-entropy of the proposed architectures near the saturation point. The untrained random network is used as a reference baseline.}}
\label{fig:ce}
\end{center}
\end{figure}

In Figure \ref{fig:ce} one can see cross-entropy of the language model (LM), VAE and VRASH architectures near the saturation point. The first untrained random network is used as a reference baseline. For the three other architectures shown in Figure \ref{fig:arcVAE}, and Figure \ref{fig:arcVRASH} the columns show the cross-entropy of the model near the point of saturation. LM and VRASH models demonstrate comparable cross-entropy with the values of $2.34$ and $2.11$ respectively.

Despite the fact that formally VRASH demonstrates only marginally better performance in comparison with the language model we claim that the results produced by VRASH are more interesting subjectively and further development of this architecture in context of music generation looks promising. Subjectively assessing the tracks produced by different algorithms we claim that the percentage of tracks with more interesting temporal and melodic structures is the highest for VRASH. \cite{Ano18} highlights artistic applications of VRASH architecture and addresses qualitative feedback from the listeners as well as professional musicians.

All three proposed architectures work relatively well and generate music that is diverse and interesting enough if the dataset for training is big enough and has high quality, however, they have certain important differences. The first general problem that occurs in many generative models is the tendency to repeat a certain note. This difficulty is more prominent for Language Model whereas VAE and, specifically VRASH tend to deal with this challenge better. 

Another issue is the macro structure of the track. Throughout the history of music a number of standard music structures were developed starting with a relatively simple structure of a song (characterized with a repetitive chorus that is divided with verses) and finishing with symphonies that comprise a number of different less sophisticated forms. Despite the fact that VAE and VRASH specifically are developed to capture macrostructures of the track they do not always provide distinct structural dynamics that characterizes a number of human-written musical tracks. However, VRASH seems to be the right way to go.

Every generative model, based on artificial neural networks has a problem of low-quality output. Currently among melodically diverse and acoustically pleasing tracks one would inevitably hear tracks with annoyingly simple recurrent patterns, off-beat sequences, obscure macrostructure etc. When faced with this problem we suggested the following workaround. Alongside with the generative VRASH-based model we proposed a set of automated filtering heuristics, that allows to obtain a pseudo-realtime non-stop music generation with very limited computational power. 

The heuristics was obtained in a straight-forward manner yet turned out to be extremely effective. Using human assessment of 1000+ tracks we have trained a classifier to predict weather a track is acoustically pleasant. Human peers were asked to evaluate tracks on a scale from 1 to 5, where 5 was the highest mark. Then we have split evaluated tracks into two categories: those that had a mark of 4 or 5 were considered acceptable, whereas the tracks that were marked with 3, 2 or 1 were to be filtered. For each track in the training dataset we have calculated the following set of theoretic informational features: 
\begin{itemize}
\item entropy of notes without octave information;
\item entropy of changes between consecutive notes without octave information;
\item entropy of notes lengths;
\item entropy of changes between consecutive notes lengths;
\item entropy of notes with octave information;
\item entropy of changes between consecutive notes with octave information;
\item minimal entropies for sliding windows that were 8, 16, 32, 64 and 128 notes long;
\item average entropy for sliding windows that were 8, 16, 32, 64 and 128 notes long;
\item coordinates of the sampling vector.
\end{itemize}

Due to the size of the dataset we were limited in our choices of methods. Table \ref{tab:fil} shows how different methods perform depending on the size of the test dataset.

\begin{table}[t!]
\small
\begin{tabular}{|r|r|r|r|r|}
\hline 
\bf  Method& \bf Share & \bf \# of bad & \bf \# of good & \bf Good tracks  \\ 
\bf  & \bf of test  & \bf tracks& \bf tracks& \bf recall on test   \\ 
\hline
 Logistic regression 	& 0.5  & 207 & 44 & 0.79  \\
 				& 0.4  & 189 & 35 & 0.81  \\
	 			& 0.3  & 162 & 24 & 0.85  \\
	 			& 0.2  & 128 & 18 & 0.86 \\ \hline
SVC with linear kernel 	& 0.5  & 238 & 64 & 0.74  \\
 					& 0.4  & 199 & 43 & 0.78  \\
	 				& 0.3  & 138 & 22 & 0.84  \\
	 				& 0.2  & 60 & 9 & 0.85  \\ \hline
SVC with rbf kernel 		& 0.5  & 229 & 53 & 0.77  \\
 					& 0.4  & 205 & 39 & 0.81  \\
	 				& 0.3  & 162 & 30 & 0.81  \\
	 				& 0.2  & 86 & 11 & 0.87  \\ \hline
\end{tabular}
\caption{\label{tab:fil} Accuracy of the filtering mechanism varies across different test sets and methods but allows to highlight up to 87\% of good tracks accurately.}
\end{table}

If there is need to speed-up the filtering one can replace the obtained classifier with a set of manually constructed empirical heuristics. Due to the fact that we are not interested in the recall of the obtained classifier (generative neural VRASH is extremely potent and robust and one faces an excessive amount of generated melodies, yet wants to filter more pleasing ones), one can make such heuristics even more strict so that 100\% accuracy is achieved.

\begin{figure}[h]
\begin{center}
\includegraphics[width=\linewidth]{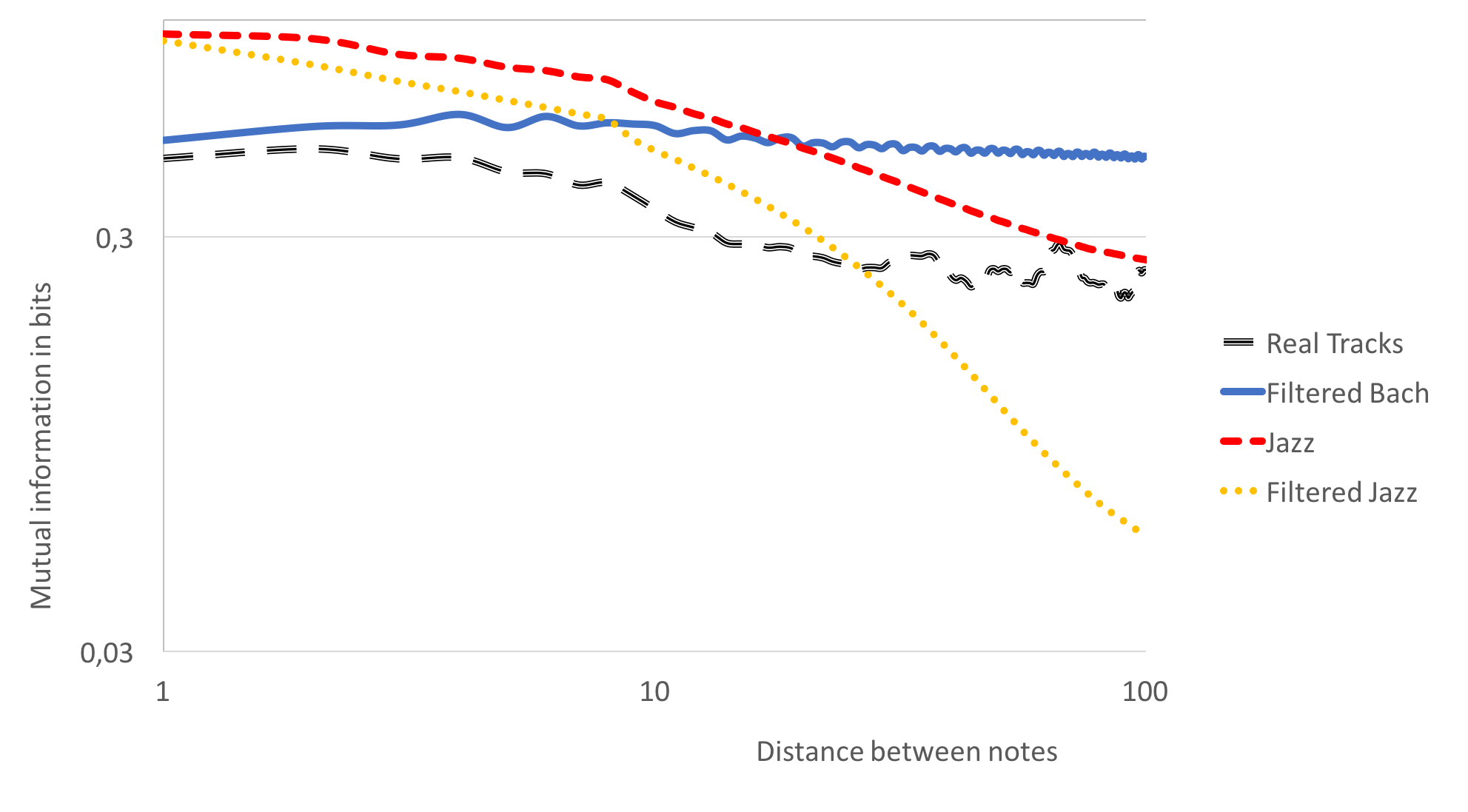} 
\caption{\emph{Mutuali information as a function of distance between two notes in real musical tracks, in VRASH-generated and automatically filtered Bach-stylized tracks, in VRASH generated jazz-stylized tracks and in in VRASH generated automatically filtered jazz-stylized tracks.}}
\label{fig:mi}
\end{center}
\end{figure}

One more way to compare generated music with real tracks is to build mutual information plots analogously to \cite{lin}. Indeed, we have earlier claimed that VRASH is designed to capture long-distance dependencies between the notes in a track. Figure \ref{fig:mi} shows how mutual information declines with distance in different types of VRASH-generated tracks. 

Looking closely on Figure \ref{fig:mi} one can notice several interesting details. First of all, Bach-stylized VRASH-generated music tend to have higher mutual information between the notes that are further apart. Similarly to real tracks mutual information declines slowly (if at all), yet its' higher values might explain the feedback that we have often got from human peers: they noticed that music was harmonious yet 'mechanistic'. Higher levels of mutual information between distant notes can partially account for that. Second, jazz-stylized VRASH generated music demonstrates a mutual information profile that is the closest to the profile of real tracks. However, as the distance between the notes gets longer, information in generated track tends to decrease faster than in real data. It also corresponds to the qualitative feedback of human peers that tend to call jazz-stylized music diverse and more human-like. Finally, filtering jazz-stylized music significantly affects the decline of mutual information between the notes. This might be to the fact that the filter was trained on Bach-stylizations. One filter that manages to provide high quality melody stream for certain style has to be retrained for another style of music on order to guarantee that it preserves the complexity needed for music to stay entertaining.





\section{Conclusion}

In this  paper we have described several architectures for monotonic music generation. We have compared Language Model, Variational Autoencoder and Variational Recurrent Autoencoder Supported by History (VRASH). This is the first application of VRASH to music generation that we know of. There are several strong advantages of this model that make it especially interesting in context of the automated music generation. First of all, it provides a good balance between the global and local structure of the track. VAE allows to focus on the macrostructure but advancing it in the way described above enables a network to generate more locally diverse and interesting patterns. Second, the proposed structure is relatively easy to implement and train. The last, but not the least, it allows to control the style of the output (through the latent representation of the input vector) and generate tracks corresponding to the given parameters. We have also proposed a simple filtering method to deal with a problem of non-consistent generative output and discussed insights into the structure of the obtained generative tracks and their comparison with actual empirical data.



\bibliographystyle{acl_natbib_nourl}
\bibliography{v2draft.bbl}

\end{document}